\documentclass[a4paper,aps,superscriptaddress,floatfix,nofootinbib,twocolumn]{revtex4-1}

\usepackage{bbold}
\usepackage{bbm}
\usepackage[pdftex]{graphicx}
\usepackage{latexsym,amsmath,verbatim,amssymb,txfonts}
\usepackage{color}
\usepackage{rotating}
\usepackage{verbatim}
\usepackage{multirow}
\usepackage[english]{babel}
\usepackage{comment}
\usepackage{hyperref}

\newcommand{\SMa}{SM1~\cite{SM1}}
\newcommand{\SMb}{SM2~\cite{SM2}}

\newcommand{\change}[1]{#1}

\begin{document}

\title{Systematic comparison between methods for the detection of influential spreaders in complex networks}

\author{\c{S}irag Erkol}
\affiliation{Center for Complex Networks and Systems Research, School
  of Informatics, Computing, and Engineering, Indiana University, Bloomington,
  Indiana 47408, USA}

\author{Claudio Castellano}
\affiliation{Istituto dei Sistemi Complessi (ISC-CNR), Via dei Taurini 19, I-00185 Roma, Italy}

\author{Filippo Radicchi}
\affiliation{Center for Complex Networks and Systems Research, School
  of Informatics, Computing, and Engineering, Indiana University, Bloomington,
  Indiana 47408, USA}
\email{filiradi@indiana.edu}

\begin{abstract}

Influence maximization is the problem of finding the set of nodes of a network
that maximizes the size of the outbreak of a 
spreading process occurring on the network.  
Solutions to this problem are important for strategic decisions in
marketing and political campaigns. The typical setting 
consists in the identification of small sets of
initial spreaders in very large networks. This setting makes the 
optimization problem computationally infeasible
for standard greedy optimization algorithms that account
simultaneously for information about network topology and
spreading dynamics, leaving space only to heuristic methods based 
on the drastic approximation of relying on the geometry of the
network alone. The literature on the subject is plenty of 
purely topological methods for the identification of 
influential spreaders in networks. However, it is unclear how far these
methods are from being optimal. Here, we perform a systematic test of the
performance of a multitude of heuristic methods for the
identification of influential spreaders. We quantify the performance
of the various methods on a corpus of $100$ real-world networks;
the corpus consists of networks small enough for the application of
greedy optimization so that results from this algorithm are used as the
baseline needed for the analysis of the performance of the other
methods on the same corpus of networks. 
We find that relatively simple network metrics, such as adaptive
degree or closeness centralities, are able to achieve
performances very close to the baseline value, thus providing
good support for the use of these metrics in large-scale problem settings.
Also, we show that a further $2-5\%$ improvement towards the baseline 
performance is 
achievable by hybrid algorithms that combine two or more topological metrics
together. This final result is validated on a small collection of large graphs
where greedy optimization is not applicable.
\end{abstract}

\maketitle


\section{Introduction}
Every day, we witness the dissemination of new pieces of information 
in social networks~\cite{ratkiewicz2011truthy, acemoglu2010spread,
  del2016spreading, 
centola2010spread, lerman2010information}. 
Few of them become widespread; the vast majority, however, 
diffuse only over a vanishing portion of the network.
Are there {\it a priori} identifiable features that 
allow for the early prediction of the outcome of a spreading process in a network?  
Many studies  have pointed out that the ``quality'' or ``attractiveness'' of the information 
might have an effect on how far it may spread
\cite{ratkiewicz2011truthy, notarmuzi2018analytical}. 
In mathematical models of information spreading, 
the notion of quality is typically
quantified in terms of the probability of spreading events along 
individual edges in the social network. However, the spreading probability
of individual edges is not the only key factor that determines
the fate of a piece of information spreading in a network. 
The nodes that act as seeds for the 
spreading process
may play a role that is more important than the actual 
probability to spread information along social contacts. 
Intuitively, if the diffusion process is seeded by central nodes, 
then the piece of information may reach large 
popularity; on the other hand, a piece of information 
originated from peripheral nodes is much less likely 
to become widespread.

The problem of selecting the best set of seed nodes 
for a spreading process in a network has been 
traditionally named as the problem of influence maximization. 
The problem is generally considered under the strong assumption
of having full and exact knowledge
of both the network topology and the spreading dynamics. 
We will adopt this line here too, 
although we remark that such an assumption is 
at least optimistic and may potentially lead \change{, if not
  satisfied, } to 
significant mistakes in the identification of the true
influential spreaders~\cite{erkol2018influence}. 
The function that is optimized in influence maximization is the
average value of 
the outbreak size. The optimization problem is solved for a 
given size of the seed set, generally much smaller than the 
network size. The problem was first formulated by 
Domingos and Richardson~\cite{domingos2001mining}, 
and later generalized by Kempe \emph{et
  al.}~\cite{kempe2003maximizing}. In particular,
Kempe \emph{et al.} showed 
that influence maximization is a NP-hard problem,
 exactly solvable for very small networks only. Also,
Kempe \emph{et al.} demonstrated that for specific models of opinion
spreading, such as the independent cascade and the
linear threshold models, the average outbreak size is a submodular
function, and thus greedy optimization algorithms allow to find, in
polynomial time, approximate solutions 
that are less than a factor $(1 - 1/e)$ 
away from the true optimum~\cite{Nemhauser1978}. 
The greedy algorithm actively uses information about the topology 
of the network and the dynamical rules of the spreading model. 
After the seminal work by Kempe \emph{et al.}, other similar
greedy techniques for 
approximating solutions to the influence maximization 
problem have been proposed~\cite{leskovec2007cost, chen2009efficient,
  goyal2011celf++, cheng2013staticgreedy}.  
As all these algorithms require knowledge of the model at the basis of 
the spreading process, often obtained through numerical simulations,
they all suffer from the limitation of being applicable to
small-medium sized networks only. We remark that
some attempts of greedy-like algorithms applicable to large networks 
have been  made~\cite{nguyen2016stop, hu2018local}. Those attempts, however,
rely on approximate estimations of the outcome of numerical
simulations, thus leading to solutions to the influence maximization
problem that are generally inferior to the solutions 
obtained with straight greedy optimization.

On large networks, like those of interest in practical applications, 
solutions to the influence maximization problem are
generally obtained via heuristic methods. The literature is full of
examples~\cite{LU20161,zhang2016identifying, lu2011leaders,
  estrada2005subgraph, chen2013identifying, de2014role,
  klemm2012measure}. 
Heuristic methods use complete information about
the network structure, but they completely neglect 
information about the dynamical model of spreading.  
They are generally much faster than greedy algorithms,
but clearly less effective. Their main limitations are two-fold.
On the one hand, heuristic methods are characterized by the 
inability to account for the combined effect 
that seeds may have in a complex spreading process, 
as the set of influential nodes is built combining the best 
individual spreaders and their influence sets may be strongly overlapping.
On the other hand, being based on purely topological properties, 
heuristic methods lack sensitivity to the features of the spreading
dynamics and the variation of the associated parameters. 
Given the wealth of heuristic methods that have been
proposed to identify influential nodes in networks, 
how different these methods are in terms of performance?
Even more important, how far is the performance of 
the best heuristic methods from optimality, 
at least the achievable optimality provided by greedy algorithms?
We realized that no clear answer 
to these fundamental questions can be found in current literature, 
and we decided to fill this gap of knowledge here.

The present paper reports on a systematic test
of $16$ heuristic methods that have been
proposed to approximate solutions to the 
influence maximization problem. Our analysis is based
on a corpus of $100$ real-world networks, and performance
of the various heuristic methods is quantified for SIR-like 
spreading processes.  Despite the various methods rely on rather 
different centrality metrics, we find that many of them 
are able to achieve  comparable performances. 
When used to select the top $5\%$ initial seeds of spreading in real-networks,
the best performing methods show levels of performance that are within 
$90\%$ from those achievable by greedy optimization, 
so that the room for potential improvement appears small. 
We show that one way to achieve better performances is
relying on hybrid methods that combine 
two or more centrality metrics together.
We validate this final result on a small set of large-scale networks.


\section{Methods}
\subsection{Networks}
In this study, we focus most of our attention on a corpus of $100$, 
undirected and unweighted, real-world networks. 
 Sizes of these networks range from $100$ to $30,000$ nodes, 
and their density varies between $0.0001$ and $0.25$.
The corpus is composed of
networks of small to medium size on purpose, as
these allow for the application of greedy optimization in the solution
of the influence maximization problem.
We consider networks from different domains. 
Specifically, our corpus of networks include $63$ social, $16$
technological, $10$ information, $8$ biological, and $3$ transportation networks. 
Details about the analyzed networks can be found in the \SMa.
In the final part of the paper, we validate some of our
findings on $9$ large real-world social and information
networks with sizes ranging from $50,000$ to 
slightly more than $1,000,000$. Details are provided in Table~\ref{table3}.

\subsection{Spreading dynamics}
We concentrate our attention on the Independent Cascade Model
(ICM)~\cite{kempe2003maximizing}. This is a very popular model in
studies focusing on the influence maximization problem. The ICM is a
simplified version of the Susceptible-Infected-Recovered (SIR)
model~\cite{pastor2015epidemic}. Nodes can be in either one of the three
states S, I, or R. At the beginning of the dynamics, all nodes start
in the S state except for those who are selected to be the initial
spreaders, which are assigned to the I state. At each step of the
model, all nodes in state I try to infect their neighbors in state S
with probability $p$; then, they recover immediately, by changing
their states from I to R. Nodes in state R never change their state and
no longer participate to the spreading dynamics.
The dynamics continue until there are no nodes left in state I. 
The size of the outbreak is calculated by counting the number of nodes 
that ended up in state R at the end of the spreading
dynamics. As the spreading from one node to another happens with
probability $p$, the model has a stochastic nature. To properly account
for the stochastic nature of the model, all our results are obtained as
average values over $50$ independent numerical 
simulations for every given initial condition.

\subsection{Methods for the selection of influential spreaders}

\begin{table}[!htb]
\begin{center}
\begin{tabular}{|l|l|l|c|r|}\hline
\textbf{Group} & \textbf{Method} &  \textbf{Abbrev.} & \textbf{Ref.} & \textbf{Complexity} \\\hline\hline
\multirow{2}{*}{Baseline} & Greedy & G & \cite{chen2009efficient} &
                                                                    cubic \\\cline{2-4}
& Random & R & - & constant \\\hline\hline
\multirow{2}{*}{Local} & Degree & D & - & linear \\\cline{2-4}
 & Adaptive Degree & AD & \cite{chen2009efficient} & linear \\\hline\hline
\multirow{8}{*}{Global} & Betweenness & B & \cite{freeman1977set} & quadratic \\\cline{2-4}
 & Closeness & C & \cite{sabidussi1966centrality} & quadratic \\\cline{2-4}
 & Eigenvector & E & \cite{bonacich1972factoring} & linear \\\cline{2-4}
 & Katz & K & \cite{katz1953new} & linear \\\cline{2-4}
 & PageRank & PR & \cite{brin1998anatomy} & linear \\\cline{2-4}
 & Non-backtracking & NB  & \cite{martin2014localization} & linear \\\cline{2-4}
 & Adaptive NB & ANB & \cite{Braunstein12368} & quadratic \\\hline\hline
\multirow{6}{*}{Intermediate} & k-shell & KS & \cite{kitsak2010identification} & linear \\\cline{2-4}
 & LocalRank &  LR & \cite{chen2012identifying} & linear \\\cline{2-4}
 & h-index & H & \cite{lu2016h} & linear \\\cline{2-4}
 & CoreHD & CD & \cite{zdeborova2016fast} & linear \\\cline{2-4}
 & Collective Influence, $\ell =1$ & CI1 & \cite{morone2015influence} & linear
  \\\cline{2-4}
& Collective Influence, $\ell = 2$ & CI2  & \cite{morone2015influence} & linear \\\cline{2-4}
& Expl. Immunization & EI & \cite{clusella2016immunization} & linear\\\hline
\end{tabular}
\end{center}
\caption{Methods for the selection of influential spreaders.
We list basic details of all the methods for the detection of influential spreaders
in complex networks that we consider in this study. Each row of the
table refers to a specific method. From left to right, 
we report the full name of the method, the abbreviation 
of the method name, the reference of the paper where the method was introduced,
and the computational complexity of the method. 
Computational complexities reported in the table are obtained under
the realistic 
assumption that methods are applied to sparse networks where the number of 
edges scales linearly with the network size.
Methods are further grouped into different categories, i.e., baseline,
local, global, and intermediate, depending on their properties.}
\label{table}
\end{table}

In total, we consider $18$ methods for the identification of
influential spreaders in networks (see Table~\ref{table}). Each method
outputs a list of nodes in a specific order from the most influential
node to the least influential node.  We use this rank to construct, in
a sequential manner, the set of the top spreaders according to a
particular method.  The various methods take as input different
type/amount of information, and make use of rather different types of
rankings.  As a consequence, the computational complexity of the
various methods may be significantly different.  For illustrative
purposes, we decided to group the $18$ methods for the selection of
influential spreaders into four main groups.

The group of baseline methods is formed by the methods greedy and random. 
The greedy algorithm is the best performing method available on the market, 
thus providing an upper bound for the performance of all other methods. 
The greedy algorithm uses all available information about network 
topology and spreading dynamics. For instance, the algorithm provides
different solutions depending on the value of the spreading
probability $p$. For the greedy method applied to the ICM, we rely on
the Chen \emph{et al.}'s \cite{chen2009efficient} algorithm, which
makes use of the mapping between ICM and bond percolation to obtain
faster results regarding the simulations of the spreading process. The
random 
method instead represents a lower bound  
for the performances of other methods. The method just outputs nodes 
of the network in random order, {\it de facto} neglecting any 
prior information regarding 
system topology and dynamics.

The remaining $16$ of the $18$ methods are purely topological methods
in the sense that they rely on heuristics that are calculated using
full knowledge of the network structure, but no information 
at all about spreading dynamics. According to these methods the
influence of a node is proportional to a network centrality metric.
Depending on the nature of the centrality metric used, 
we classify the topological methods into three groups. 

First, methods that use local topological information, in the sense
that values of the 
centrality metric associated to every node are computed using
information about 
their nearest neighbors only.  For example, degree centrality, which
consists 
of counting the number of neighbors of a node, 
belongs to this category. A variant of the degree method, called 
adaptive degree method, which was proposed by Chen \emph{et al.}~\cite{chen2009efficient} 
is classified as a local method too. 

Second, methods that are based on global centrality metrics whose
computation, at the level of the individual nodes, requires complete knowledge
about the whole network structure. This group consists of methods relying on 
betweenness~\cite{freeman1977set},
closeness~\cite{sabidussi1966centrality},
eigenvector~\cite{bonacich1972factoring}, Katz~\cite{katz1953new}, 
non-backtracking~\cite{martin2014localization,
  radicchi2016leveraging}, 
and pagerank~\cite{brin1998anatomy} centralities. 
As a part of this group we also considered the method based on an 
adaptive variant of the non-backtracking centrality~\cite{Braunstein12368}. 

Finally, we consider several methods that rely on intermediate
topological information (e.g., nearest neighbors, next-nearest
neighbors) for the 
computation of node centrality metrics. This group consists of the
methods that rely on the metrics
k-shell~\cite{kitsak2010identification},
localrank~\cite{chen2012identifying}, and h-index~\cite{lu2016h}. We
classify in the intermediate group also methods that are based on
collective influence~\cite{morone2015influence},
coreHD~\cite{zdeborova2016fast}, and explosive immunization
score~\cite{clusella2016immunization}. These are methods introduced
with the goal of approximating solutions to the optimal percolation
problem~\cite{morone2015influence}, an optimization problem that has
similarities with, but is
different from the one considered in influence 
maximization~\cite{radicchi2017fundamental}. 
We stress that we consider two variations of the $CI$
method. Specifically,
we consider $CI1$ and $CI2$, where
 the numerical value indicates the value of the parameter that
defines the centrality metric~\cite{morone2015influence}.

\subsection{Evaluating the performance of methods for the selection
of influential spreaders}

\begin{figure}[!htb]
\includegraphics[width=0.45\textwidth]{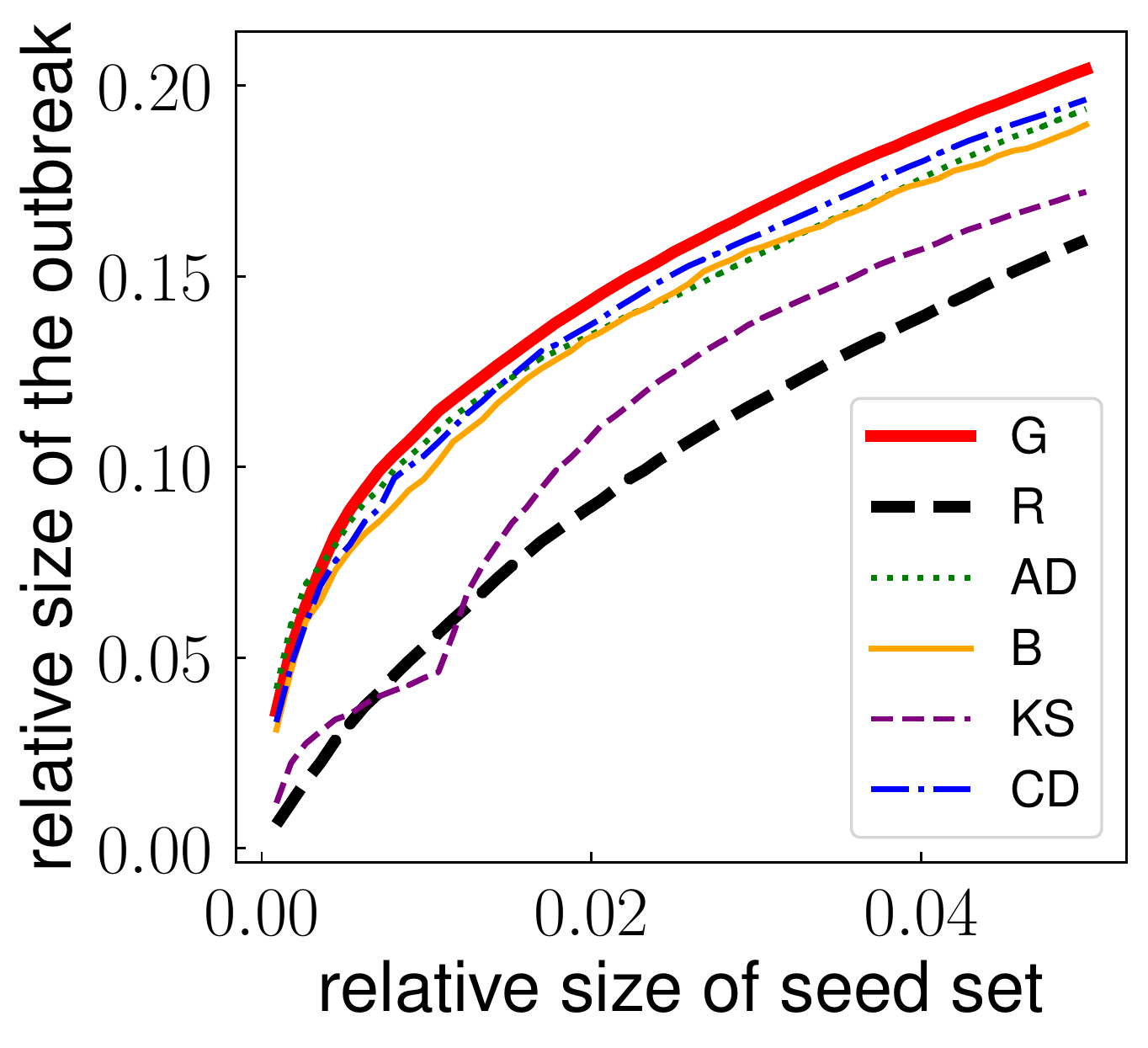}
\caption{Relative size of the outbreak as a function of the relative size of 
the seed set for the email communication network of Ref.~\cite{guimera2003self}.
To obtain relative values, we divide outbreak size and seed set size by the
total number of nodes in the network. Relative measures allow for an immediate
comparison across networks with different sizes.
We compare the performance of different methods for the selection of influential nodes.
Outbreak size is calculated for ICM dynamics at critical threshold
$p_c = 0.056$.
To avoid \change{ overcrowding}, we display results
only for a subset of the methods considered in the paper.
}
\label{fig:single1p}
\end{figure}

\change{Potentially all selection methods described above 
are subjected to statistical fluctuations in the sense 
that they may generate  a different ranking for the nodes
at each run. This is due to the presence of ties in the ranking of
nodes, and the fact that we break ties by randomly selecting nodes
with the same rank position.}
To account for \change{statistical fluctuations}, we apply every 
method $R = 10$ independent times 
to generate $R$ rankings for the nodes.
We consider each of these rankings to sequentially 
construct sets of top spreaders. Specifically,
we indicate as $\mathcal{S}_{m}^{(t,r)}$ the set of top $t \, N$ 
spreaders identified by method $m$ in instance $r$ of the
method and for a given network with $N$ nodes. 
For every set $\mathcal{S}_{m}^{(t,r)}$, we 
run $50$ different times the ICM model,
and measure the average value of the
outbreak size $O[\mathcal{S}_{m}^{(t,r)}]$.
We then repeat the operation for every instance 
$r$ of the method, and take the average 
over the $R$ potentially different sets, namely
\begin{equation}
V_m^{(t)}  = \frac{1}{R} \sum_{r=1}^R \,
O[\mathcal{S}_{m}^{(t,r)}] \; . 
\label{eq:vm}
\end{equation}
Fig.~\ref{fig:single1p} displays how 
the relative size of the outbreak $V_m^{(t)} / N$ grows 
as function of the  relative seed set size $t $ for some of the methods for the
identification of top spreaders considered in this paper.
\change{Given the amount of simulations performed, 
the standard error associated with the average value of the
  outbreak size of Eq.~(\ref{eq:vm}) is always very small. We therefore
  neglect it in all the considerations and analyses below.}
Fig.~\ref{fig:single1p} clearly shows that the greedy 
and random algorithms are good baselines
for the performances of the other methods. 
For instance, the greedy algorithm outperforms all other
methods. This result is confirmed across the entire corpus of
networks we analyzed in this paper (see \SMa ~and \SMb).
\change{In a few networks, some heuristic methods
are able to slightly outperform the greedy algorithm. This seems to
happen only in the case of relatively small networks, composed of
hundreds or less nodes.}
Similarly, all methods perform better than the random selection method, 
although there are quite a few cases where randomly selecting seeds 
perform as well as selecting seeds according to some topological heuristic.

As a measure for the performance of method $m$
in the identification of the top $T \, N$  
influential spreaders of a given network, we evaluate the area under 
the curves of Fig.~\ref{fig:single1p} up to a 
pre-imposed $T$ value 

\begin{equation}
q_m^{(T)}  = \frac{1}{N} \, \int_{0}^{T}  dt \, V_m^{(t)} \; .
\label{eq:outbreak}
\end{equation}

 As the size of set
of top spreaders are linearly dependent from the
size of the network $N$, we can easily aggregate results
obtained over the entire corpus of real-world networks at our disposal.
Specifically, results in the main paper are obtained for
$T=0.05$. We report results for $T=0.1$
in the \SMb. No significant differences between the two cases are apparent.  
As some of the methods considered in the paper are 
characterized by large computational 
complexity (see Table~\ref{table}), we couldn't consider $T > 0.1$.
We note, however, that studying the performance of methods for
the identification of influential spreaders has a meaning 
only for small $T$ values, given that in practical applications 
the seeding is generally performed on a vanishing portion of the
system.
Also, we test the validity of all results using $V_m^{(T)}$ as a main 
metric of performance, instead of its integral of
Eq.~(\ref{eq:outbreak}). Results are reported in the \SMb. 
No significant changes with respect to the results presented here in the
main paper are apparent.

\begin{figure}[!htb]
\includegraphics[width=0.45\textwidth]{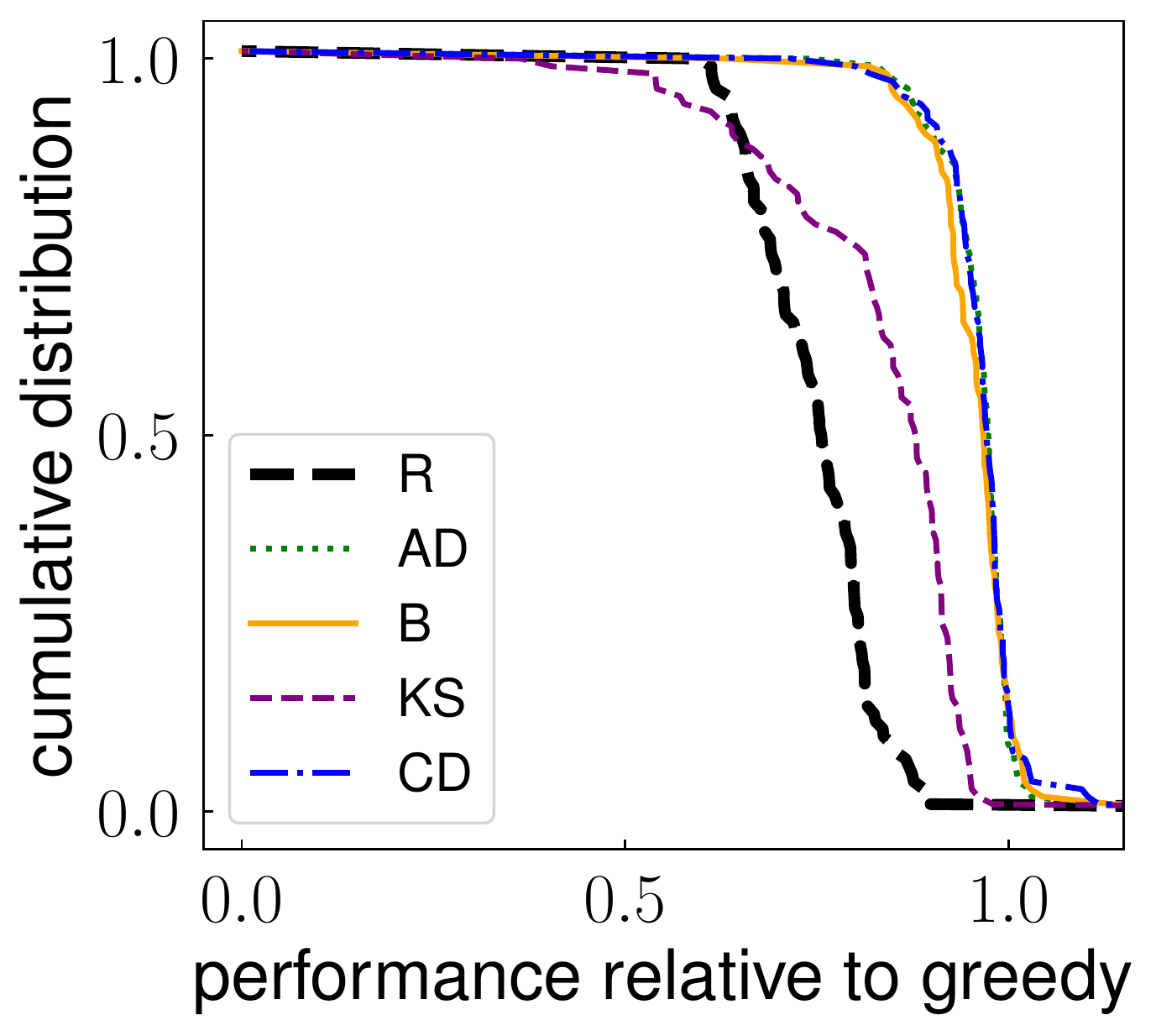}
\caption{
Cumulative distribution of the relative performance $g_m^{(T)}$
(for $T=0.05$) obtained by using a method for the identification 
of influential spreaders different from the greedy algorithm.
The metric of relative performance  is defined in Eq.~\ref{eq:gain}.
The distribution is obtained considering all networks in our dataset. 
For every network, the outbreak size is calculated for ICM dynamics 
at critical threshold $p_c$.
See details in the \SMa.  
\change{
To avoid overcrowding, we display results
only for the same subset of the methods as already considered in Fig.~\ref{fig:single1p}.}
}
\label{fig:ccdf1p}
\end{figure}

As the greedy algorithm
provides an upper bound for the performance 
of the other methods, we use it as a term of comparison 
for all other methods in our systematic analysis. 
We consider two main metrics of performance.
The first measure is based on a comparison between
the outbreak size obtainable by a method compared to the one 
obtained using the greedy identification method.  Specifically, 
given a network, we first compute
\begin{equation}
g_m^{(T)} = \frac{
q_m^{(T)}}{q_{G}^{(T)}} \; ,
\label{eq:gain}
\end{equation}
where
we used the abbreviation $q_{G}^{(T)}$ to indicate
the expression of Eq.~(\ref{eq:outbreak}) for the greedy algorithm,
i.e., $m=G$.  Then, we 
evaluate the performance relative to greedy for all networks in our dataset, and 
summarize the results in Fig.~\ref{fig:ccdf1p} where we display 
the cumulative distribution of this quantity for some of the methods.
To obtain a single number for the performance of the method over the
entire corpus of networks,  we define the 
overall performance $\langle g_{m}^{(T)} \rangle$
given by the average value of the metric defined in Eq.~(\ref{eq:gain})
over all real networks in the dataset.
\change{We remark that statistical errors associated to the metrics 
of Eqs.~(\ref{eq:vm}), ~(\ref{eq:outbreak}) and
~(\ref{eq:gain}) are negligible given the large number of
independent numerical simulations used to determine 
their average values. A similar statement, however, doesn't hold 
for the overall performance $\langle g_{m}^{(T)} \rangle$ 
due to the relatively small size of the corpus
  of networks analyzed. In the following, we associate the
  standard error of the mean to any  
estimate of the average value $\langle g_{m}^{(T)} \rangle$ 
obtained on samples of real-world networks.}

\begin{figure}[!htb]
\includegraphics[width=0.45\textwidth]{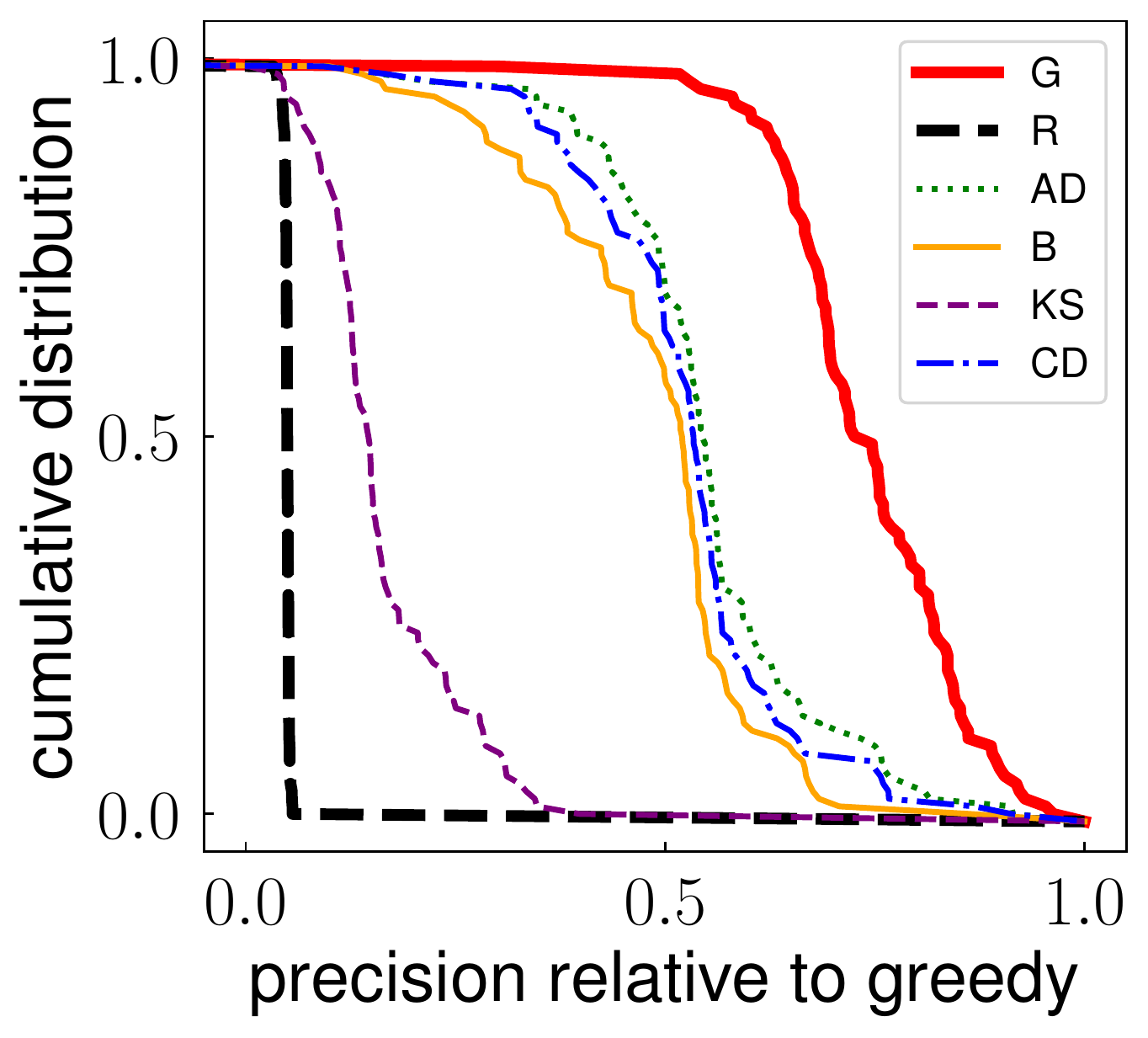}
\caption{Cumulative distribution of the precision metric $r_m^{(T)}$
defined in Eq.~(\ref{eq:precision}) for $T=0.05$. The distribution is 
obtained considering all networks in our dataset. 
Results for the greedy algorithm used in the comparison are 
those obtained for
ICM dynamics at critical threshold $p_c$.
See details in the \SMa.  
\change{
To avoid overcrowding, we display results
only for the same subset of the methods as already considered in Fig.~\ref{fig:single1p}.}
}
\label{fig:prec1p}
\end{figure}

The second metric of performance instead
neglects the size of the outbreak, 
and focuses only on the identity
of the nodes identified by the method $m$. 
For the actual solution of the problem of influence maximization,
this second metric is clearly much less important than the one previously
considered. However, the metric can tell us something more
about the topological properties of the set of top spreaders in
networks.  Given a network, we evaluate 
the frequency $f_m^{(T,i)}$ of every node $i$ 
to be in the set of top $T \, N$ spreaders according
to method $m$  
over $R = 10$ runs of the algorithm. We then 
compute the precision of the method relative to the greedy algorithm as
\begin{equation}
r_m^{(T)} = \frac{1}{T \, N} \, \sum_{i=1}^N \, f_m^{(T,i)} \, f_G^{(T,i)}  \; .
\label{eq:precision}
\end{equation}
We note that
Eq.~(\ref{eq:precision})
can be used to measure the self-consistency of the 
greedy method by setting $m=G$.
The cumulative distribution of the precision metric defined in
Eq.~(\ref{eq:precision}) across the entire network dataset is
displayed in Fig.~\ref{fig:prec1p}. The plot shows high level of precision
between some methods and the greedy algorithm. The random
selection method generates a distribution well peaked around
the value $T$.  We characterize the generic method $m$ with a metric of overall
precision  $\langle r_m^{(T)} \rangle$  
as the average value of the precision 
defined in Eq.~(\ref{eq:precision}) over the entire corpus of real
networks.
\change{Statistical errors associated to measure of $\langle r_m^{(T)}
  \rangle$  are quantified in terms of standard error of the mean.}
The value of $\langle r_m^{(T)}
  \rangle$ tells us how much the
method $m$ is similar to the baseline provided by the greedy algorithm
in the identification of the top spreaders across the entire corpus of
networks at our disposal.

\begin{figure*}[!htb]
\includegraphics[width=0.95\textwidth]{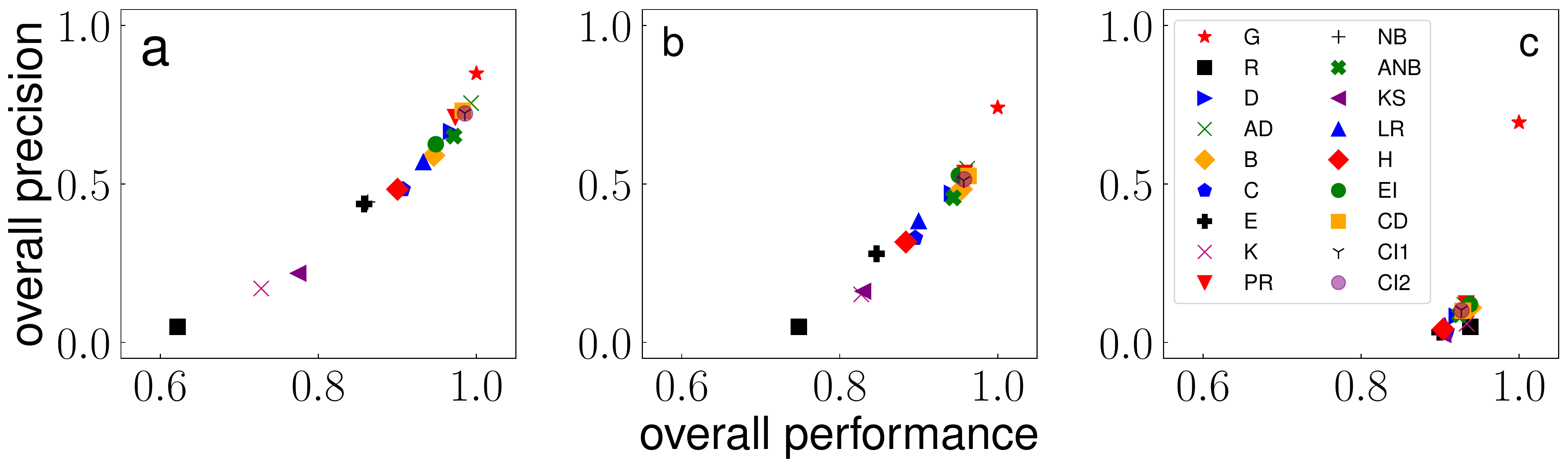}
\caption{Performance and precision of methods for the identification of influential spreaders in real networks. Results are based on the 
systematic analysis of $100$ real-world networks. 
For each network, we
first evaluate the critical value of the spreading probability $p_c$
for ICM dynamics. Then, we consider 
the analysis for three distinct phases of spreading: (a) $p
= p_c/2$,  (b) $p = p_c$, (c) $p = 2p_c$. Each point 
in the various panels corresponds to one method. 
Every method is used to identify the top $T \, N$, with 
$T= 0.05$, spreaders in the networks.
For clarity of the figure, methods are identified 
by the same abbreviations as those defined in 
Table~\ref{table}. Methods are characterized by the metrics of
performance defined in the paper. Both these  metrics
relate the performance of a generic method $m$  to the
one of the greedy algorithm. Overall performance $\langle g_m \rangle$ is a metric 
of performance that relies on the size of the outbreak 
associated with the set of influential spreaders identified by the
method compared to the typical outbreak obtained with the greedy
algorithm. Overall precision $\langle r_m \rangle$ instead quantifies the overlap between
the sets of spreaders identified by a method and those identified by
the greedy algorithm.  \change{Error bars (not shown) quantifying the 
standard errors of the mean associated
  with the numerical estimates of
$\langle g_m \rangle$ and $\langle r_m
\rangle$ are of the same size as of the symbols used in the
visualization.}
}
\label{fig:svp1p}
\end{figure*}

\begin{figure}[!htb]
\includegraphics[width=0.45\textwidth]{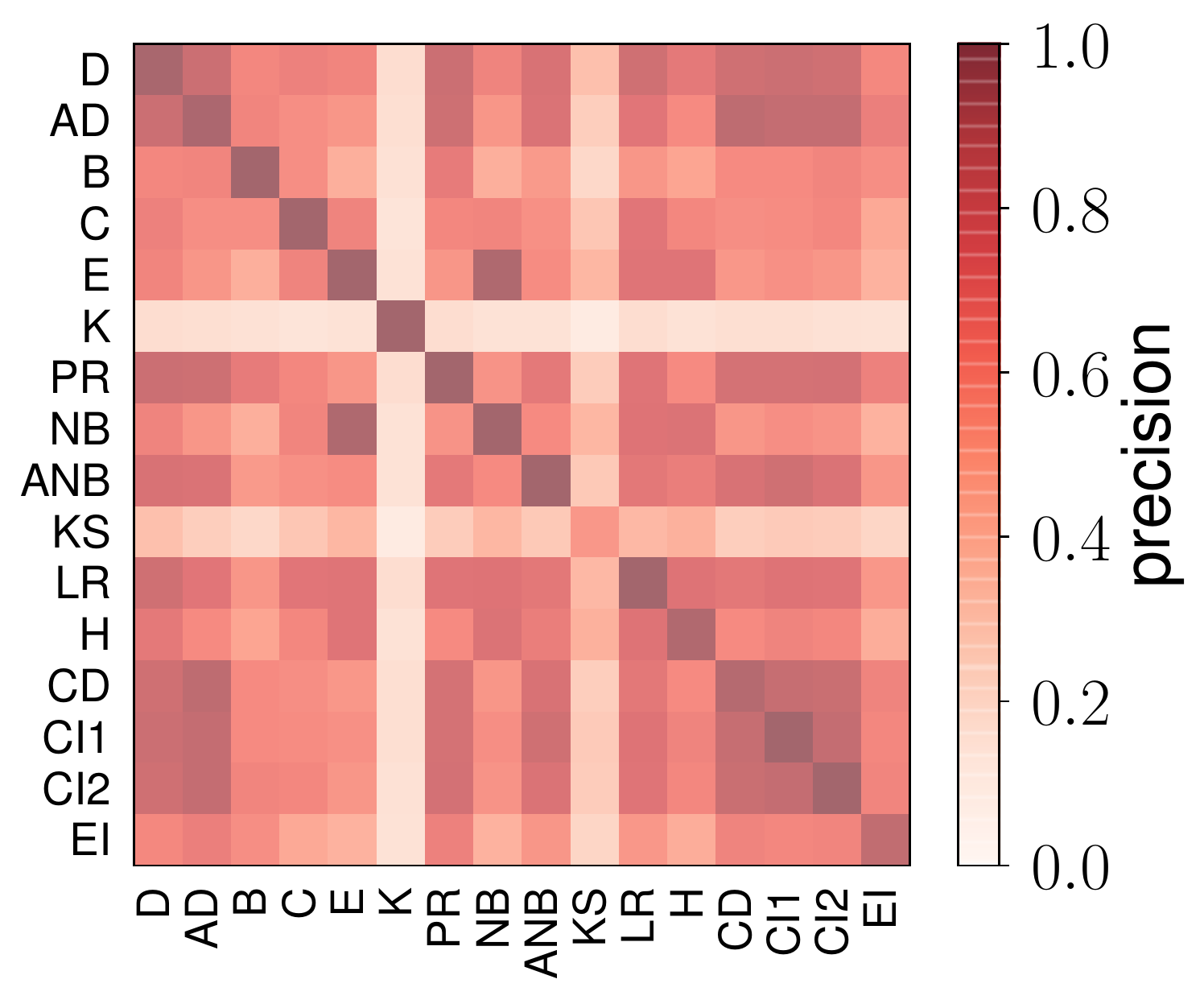}
\caption{Pairwise comparison among methods for the identification of
  influential spreaders. For every pair of methods $m_1$ and $m_2$, 
we evaluated the overlap $r_{m_1, m_2}^{(T)}$ among the two sets of top $T \, N$ influential
spreaders found by the methods in the network using  a
precision metric similar to the one of Eq.~(\ref{eq:precision}), 
i.e., $r_{m_1, m_2}^{(T)} = \frac{1}{TN} \, \sum_{i=1}^N \,
f_{m_1}^{(T,i)} \, f_{m_2}^{(T,i)}  $. We then estimated the average
value of the precision over the entire corpus of real networks at our disposal.
In the figure, dark colors corresponds to high values of precision;
low precision values are represented with light colors. Acronyms of
the methods are defined in Table~\ref{table}. Methods are listed in
the table according to the same order as they appear in Table~\ref{table}.
}
\label{fig:precision}
\end{figure}


\section{Results}

\subsection{Individual methods}

Armed with the metrics defined in the section above,
we test the various methods 
for the identification of influential spreaders for ICM 
dynamics over the entire corpus of real networks at our disposal.
We remark that both the identity and performance of the true set
of influential spreaders may be dependent on the 
actual value of the spreading probability $p$ in the ICM model, so that
the performance of the various seed selection 
methods needs to be evaluated at different values 
of the spreading probability $p$. For instance, 
for the extreme cases $p=0$ and $p=1$, 
predictions are trivial in the sense that all methods
have exactly the same performance in terms of outbreak size.
The prediction of methods performance 
is instead non trivial when the
uncertainty of the spreading outcome is maximal. For this reason, we
focus our attention on ICM dynamics around the critical threshold $p
= p_c$.  To perform the analysis, we first evaluate the critical threshold
values $p_c$ for every network in the database. Specifically, we rely
on mapping between bond percolation and the ICM, and we apply the
Newman-Ziff algorithm to evaluate $p_c$~\cite{newman2000efficient, radicchi2015predicting}.
$p_c$ values for the various networks
are reported in the \SMa.
We then consider ICM dynamics for three distinct values of $p$:
(i) subcritical regime at $p = p_c/2$;
(ii) critical regime at $p=p_c$;
(iii) supercritical regime at $p=2p_c$.

Results of our analysis are summarized in Fig.~\ref{fig:svp1p}.  Every
method is used to identify the set of top $T \, N$ nodes in the
networks, with $T= 0.05$. 
 In the figure, we represent results for each
method $m$ in the plane $(\langle g_m \rangle, \langle r_m
\rangle)$. \change{Numerical values of $\langle g_m \rangle$ and $\langle r_m
\rangle$, as well as their associated statistical errors, are reported
in \SMb.}
 Please note that we dropped the suffix $T$ to simplify the
notation.  We remark that the performance of every method $m$ is
measured in relation to the performance of the greedy method, i.e.,
$m=G$. By definition, we have $\langle g_G\rangle = 1$; we find
instead that the self-consistency score is $\langle r_G \rangle < 1$
meaning that optimal sets identified by the greedy algorithm have some
degree of variability. \change{Such a variability seems due to the
existence of (quasi)degenerate solutions to the
  influence maximization problem, i.e., different seed sets
  corresponding to similar outbreak sizes. The presence of statistical
  fluctuations in the numerical estimates of the outbreak size may
 be an additional confounding factor that exacerbates the degeneracy of
 greedy solutions.}
 An interesting finding is the absence of a
strong dependence of $\langle r_G \rangle$ from the dynamical regimes
of the ICM. The other important reference point in the plane is given
by the random method ($m=R$). By definition, we have that $\langle r_R
\rangle \simeq T = 0.05$.  $\langle g_R \rangle$ values instead
strongly depend on the dynamical regime.

In the subcritical regime (see Fig.~\ref{fig:svp1p}a), the two metrics
$\langle g_m\rangle$ and $\langle r_m \rangle$ are tightly related one
to the other. Adaptive degree ($m= AD$) outperforms all other methods
in both metrics. Other methods that perform very well are those based
on algorithms relying on the Degree ($m=D$), Adaptive Non-Backtracking
($m=ANB$) and PageRank ($m=PR$) centralities, as well as those based
on the CoreHD ($m = CD$) and Collective Influence ($m = CI$)
algorithms. 
Similar considerations apply to the critical regime (Fig.~\ref{fig:svp1p}b).  
The most significant change with
respect to the subcritical regime is a slight decrease of range of
values for the performance metric of the algorithms.  In the
supercritical regime (Fig.~\ref{fig:svp1p}c), there is no longer a
proper distinction between the various methods in terms of
performance.

A remarkable feature emerging from Fig.~\ref{fig:svp1p} is that the
overall performance is rather high. For most of the methods values are
above 0.9 for all values of $p$, and even random selection
provides a performance always larger than 0.6.
This observation somehow helps to properly weigh the importance of
greedy algorithms for influence maximization: while their solutions
are guaranteed to be not too far from the true optimum, their 
performance can be almost achieved by simple and much more easily
implemented purely topological methods.

The similarity in the performance between the various methods can be
deduced by a straight pair-wise comparison between the sets of
top influential nodes identified by the various methods across the
entire corpus of real networks at our disposal. The results of this
analysis are summarized in 
Fig.~\ref{fig:precision}. Top-performing methods provide sets
of influential nodes very similar to each other; methods with low
performance instead generally identify influential nodes that are
rarely selected by any other method. 

In the \SMb, we repeat the same exercise by computing the performance 
scores restricted to different subsets of the whole corpus of networks.
The subsets correspond to networks from the same domain (e.g., social,
technological, transportation); we do not 
find any significant change in the main outcome of the analysis. 

We further consider 
artificial networks created with the Barabasi-Albert (BA)
model~\cite{barabasi1999emergence}. Results are very similar to those
obtained on real-world networks (see \SMb). In summary, it seems that the main
results of the paper are unchanged by the nature/type of the network
substrate where spreading is occurring.

\subsection{Hybrid methods}

\begin{table}[!htb]
\begin{center}
\centering

\begin{tabular}{|c|c|c|c|c|}\hline
Method & Features & Subcrit. & Critical & Supercrit.
  \\\hline \hline

\multirow{3}{*}{AD} & $c_{AD}$ & 1.000 & 1.000
                                             & 1.000  \\ \cline{2-5}
& $\langle g_m \rangle$ & 0.993 & 0.961 & 0.931 \\  \cline{2-5}
& $\langle r_m \rangle$ & 0.755  & 0.548 & 0.119 \\ \hline\hline

\multirow{3}{*}{CD} & $c_{CD}$ & 1.000 & 1.000
                                             & 1.000 \\ \cline{2-5}
& $\langle g_m \rangle$ & 0.983 & 0.963 & 0.929 \\  \cline{2-5}
& $\langle r_m \rangle$ & 0.730  & 0.525 & 0.100 \\ \hline\hline

\multirow{3}{*}{B} & $c_{B}$ & 1.000 & 1.000
                                             & 1.000 \\ \cline{2-5}
& $\langle g_m \rangle$ & 0.946 & 0.954 & 0.938 \\  \cline{2-5}
& $\langle r_m \rangle$ & 0.590  & 0.483 & 0.110 \\ \hline\hline

\multirow{4}{*}{\shortstack{AD,B}} & $c_{AD}$ & 0.718 & 0.590 & 0.023 \\\cline{2-5}
& $c_{B}$ & -0.027 & 0.046 & 0.069 \\\cline{2-5}
& $\langle g_m \rangle$ & 0.987 & 0.964 & 0.936 \\\cline{2-5}
& $\langle r_m \rangle$ & 0.755 & 0.551 & 0.116 \\\hline \hline

\multirow{5}{*}{\shortstack{AD,PR,LR}} & $c_{AD}$ & 1.189 & 1.044 & 0.115 \\\cline{2-5}
& $c_{PR}$ & -0.266 & 0.145 & 0.772 \\\cline{2-5}
& $c_{LR}$ & -0.336 & -0.632 & -0.771 \\\cline{2-5}
& $\langle g_m \rangle$ & 0.991 & 0.980 & 0.971 \\\cline{2-5}
& $\langle r_m \rangle$ & 0.806 & 0.616 & 0.300 \\\hline \hline

\multirow{5}{*}{\shortstack{PR,LR,CD}} & $c_{PR}$ &  0.006 & 0.386 & 0.803 \\\cline{2-5}
& $c_{LR}$ & -0.419 & -0.702 & -0.771 \\\cline{2-5}
& $c_{CD}$ & 1.028 & 0.898 & 0.088 \\\cline{2-5}
& $\langle g_m \rangle$ & 0.985 & 0.979 & 0.971 \\\cline{2-5}
& $\langle r_m \rangle$ & 0.784 & 0.597 & 0.293 \\\hline \hline

\multirow{5}{*}{\shortstack{AD,B,LR}} & $c_{AD}$ & 1.096 & 1.047 & 0.343 \\\cline{2-5}
& $c_{B}$ & -0.010 & 0.067 & 0.083 \\\cline{2-5}
& $c_{LR}$ & -0.466 & -0.565 & -0.395 \\\cline{2-5}
& $\langle g_m \rangle$ & 0.993 & 0.976 & 0.952 \\\cline{2-5}
& $\langle r_m \rangle$ & 0.810 & 0.625 & 0.220 \\\hline \hline

\multirow{5}{*}{\shortstack{PR,LR,EI}} & $c_{PR}$ & 0.304 & 0.583 & 0.740 \\\cline{2-5}
& $c_{LR}$ & 0.101 & -0.251 & -0.733 \\\cline{2-5}
& $c_{EI}$ & 0.235 & 0.277 & 0.121 \\\cline{2-5}
& $\langle g_m \rangle$ & 0.973 & 0.964 & 0.970 \\\cline{2-5}
& $\langle r_m \rangle$ & 0.698 & 0.589 & 0.304 \\\hline

\end{tabular}
\end{center}
\caption{
Hybrid methods for the identification of influential
spreaders in networks. The table is organized in 
various blocks, each corresponding to a
specific method. For every method $m$, either 
individual or hybrid, we report performance values for the three
different dynamical regimes in terms of overall performance $\langle g_m
\rangle$
and overall precision $\langle r_m \rangle$. The top three blocks 
correspond to the best individual methods in the three regimes
according to overall performance metric. The remaining blocks are for 
hybrid methods. In each block, the first rows report values
of the coefficient $c_m$ of the individual method $m$ in the definition
of the hybrid method. We report the averages
for the coefficient values over $1,000$ iterations of the learning
algorithm. The bottom two rows in each block correspond
instead to the values of the performance metrics.
Errors associated with all these measures are always smaller
than $0.001$, and they are omitted from the table for clarity.
}
\label{table2}
\end{table}


\begin{table*}[!htb]
\begin{center}
\centering
\begin{tabular}{|l|r|r|r|r|r||c|c|c|}
\cline{7-9}
\multicolumn{6}{r|}{} & \multicolumn{3}{c|}{$\langle g_\mathcal{H}
                        \rangle / \langle g_{AD} \rangle$} \\
\hline
Network & $N$ & $E$ & $p_c$ & Ref. & url & Subcrit. & Critical
  & Supercrit. \\\hline
\hline
Slashdot & 51,083 & 116,573 & 0.0262 & \cite{gomez2008statistical, konect} & \href{http://konect.uni-koblenz.de/networks/slashdot-threads}{url} & 1.003 & 1.017 & 1.062 \\\hline
Gnutella, Aug. 31, 2002 & 62,561 & 147,878 & 0.0956 & \cite{ripeanu2002mapping, leskovec2007graph} & \href{http://snap.stanford.edu/data/p2p-Gnutella31.html}{url} & 1.009 & 1.040 & 1.039 \\\hline
Epinions & 75,877 & 405,739 & 0.0062 & \cite{richardson2003trust, konect} & \href{http://konect.uni-koblenz.de/networks/soc-Epinions1}{url} & 1.012 & 1.057 & 1.130 \\\hline
Flickr & 105,722 & 2,316,668 & 0.0142 & \cite{McAuley2012, konect} & \href{http://konect.uni-koblenz.de/networks/flickrEdges}{url} & 1.007 & 1.082 & 1.242 \\\hline
Gowalla & 196,591 & 950,327 & 0.0073 & \cite{cho2011friendship, konect} & \href{http://konect.uni-koblenz.de/networks/loc-gowalla_edges}{url} & 1.011 & 1.024 & 1.066 \\\hline
EU email & 224,832 & 339,925 & 0.0119 & \cite{leskovec2007graph, konect} & \href{http://konect.uni-koblenz.de/networks/email-EuAll}{url} & 1.002 & 1.009 & 0.923 \\\hline
Web Stanford & 255,265 & 1,941,926 & 0.0598 & \cite{leskovec2009community} & \href{http://snap.stanford.edu/data/web-Stanford.html}{url} & 1.009 & 1.031 & 1.035 \\\hline
Amazon, Mar. 2, 2003 & 262,111 & 899,792 & 0.0940 & \cite{leskovec2007dynamics} & \href{http://snap.stanford.edu/data/amazon0302.html}{url} & 1.008 & 1.025 & 0.994 \\\hline
YouTube friend. net. & 1,134,890 & 2,987,624 & 0.0063 &
                                                    \cite{leskovec2012,
                                                        konect} &
                                                                  \href{http://konect.uni-koblenz.de/networks/com-youtube}{url}
                                   & 1.004 & 1.013 & 0.952 \\\hline \hline
\multicolumn{6}{r|}{\it{Average on large networks}} & 1.007 $\pm$ 0.001&
                                                                       1.033
                                                                       $\pm$
                                                                       0.007
              & 1.050 $\pm$ 0.030 \\\cline{7-9}
\multicolumn{6}{r|}{\it{Average on the corpus of 100 networks}} &
                                                                  1.001
                                                                  $\pm$
                                                                  0.002
        & 1.021 $\pm$ 0.003 & 1.043 $\pm$ 0.005 \\\cline{7-9}

\end{tabular}
\end{center}
\caption{
Identification of influential spreaders in large networks.
We compare the performance of the hybrid method $\mathcal{H}$ = \{AD,PR,LR\} with the
individual method AD. For the hybrid method, we use the values of the
coefficients reported in Table~\ref{table2}. From left to right, we
report the name of the network, number of nodes in the giant
component, number of edges in the giant component, critical value
$p_c$ of
the spreading probability, references to studies where the network was
first analyzed, url where network data were downloaded, value of the
ratio $\langle g_\mathcal{H} \rangle / \langle g_{AD} \rangle$
between the performance metric of the hybrid method $\mathcal{H}$ =
\{AD,PR,LR\} 
and the one of the individual method AD for the subcritical, critical
and supercritical regimes. The bottom two lines in the table report,
for each dynamical regime,
average values \change{ and standard errors of the mean } 
for the ratios $\langle g_\mathcal{H} \rangle / \langle g_{AD} \rangle$ 
over the set of large networks and over the corpus of 100 networks
considered in the rest of the paper. 
}
\label{table3}
\end{table*}


In this section, we report on the performance of hybrid methods
for the identification of top spreaders in the network obtained from
linear combinations of the individual methods considered so far.
Specifically, we first select a certain number of individual methods
to form a hybrid 
method $\mathcal{H} = \{m_1, m_2, \ldots, m_{|\mathcal{H} |}\}$.
We associate to every node $i$ in a given network a score
$s_i^{(\mathcal{H} )}$
that is a linear combination of the scores associated with individual 
methods, namely
\begin{equation}
s_{\mathcal{H}}^{(i)} = \sum_{m \in \mathcal{H} } \, c_m \,  s_{m}^{(i)}
\; .
\label{eq:hybrid}
\end{equation}
In Eq.~(\ref{eq:hybrid}), $s_{m}^{(i)}$ is the normalized score of node $i$ in
the network according to the topological
metric used by method $m$.   
 The normalization (L$^2$-norm)
has the purpose of making scores of comparable 
magnitude across methods.
The best estimates of the linear coefficients $c_m$ are
then obtained using information from the greedy
algorithm. We use  linear regression to
find the best linear fit between $s_{\mathcal{H}}^{(i)}$ and 
$f_G^{(T , i)}$, i.e., the probability that node $i$ is identified by
the greedy algorithm in the set of top $T \, N$ influential nodes in
the network. 
Best estimates of the coefficients are obtained relying on
a training set composed of $80\%$ of networks randomly
chosen out of the corpus of real networks at
our disposal. We then test the hybrid method
$\mathcal{H}$  on the remaining $20\%$ of the corpus, where
we measure overall performance and overall precision. We replicate
the entire procedure 
$1,000$
times to quantify 
uncertainty associated with both the best estimates of the linear
coefficients as well as the measured values of the performance metrics.

We consider  several hybrid methods consisting in the combination of
two and three individual centrality metrics. 
In general, we combine together
centrality methods that differ on the basis of their classification in local,
global and intermediate methods (see Table~\ref{table}). Results for 
some hybrid methods are reported in Table~\ref{table2}.
Several remarks are in order.
First, with respect to the case of individual methods, 
there is an increase in the measured values of the
overall precision $\langle r_m \rangle$. This tells us that the
coefficients learned from the training set can be meaningfully used
on other networks to mimic greedy optimization in terms of topological
features
only. The overall performance $\langle g_m
\rangle$ of hybrid methods
increases too; improvements beat even by $2-5\%$ the best 
individual methods.
Second, when similar individual methods are combined together into an
hybrid method, 
one of the two gets the biggest part of the weight compared to the
other.
For example, the hybrid method $\mathcal{H}  = \{AD, B\} $
learned from data is almost
a pure AD method in both the subcritical and critical regimes.
Third, the
coefficients of the linear combination of Eq.~(\ref{eq:hybrid})
can also be negative. For example, for the hybrid method
$\mathcal{H}  = \{AD, PR, LR\}$ in the critical
regime, $c_{LR} < 0$. Thanks to this fact, the method outperforms in both the critical and subcritical
regimes all other methods
considered in this paper.
\change{We stress that the finding $c_{LR} < 0$   doesn't mean that LR
  centrality is anticorrelated with node  influence. $c_{LR} <
  0$, in fact, is observed only when LR is used in
  combination with other metrics. Indeed, LR centrality is 
positively correlated with node
  influence when LR is used as the only method for the identification of
  spreaders, as Figure~\ref{fig:svp1p} clearly shows.}

To validate the use of hybrid methods for the identification of
influential spreaders, we apply the top-performing hybrid method
$\mathcal{H} = \{AD, PR, LR\} $ to large social and information
networks. Results are reported in Table~\ref{table3}.  These networks
are too big for the application of greedy optimization, thus the
performance of the hybrid method is compared to the one of the method AD by
taking the ratio $\langle g_{\mathcal{H}} \rangle / \langle g_{AD}
\rangle $.  Please note that AD is one of the best individual methods
for the identification of influential spreaders according to our
analysis on the corpus of small/medium networks.  When applying the
hybrid method to large networks, we use the same values of the linear
coefficients learned from small/medium networks and listed in
Table~\ref{table2}. Overall, we see that the hybrid method generates
improvements in the detection of influential spreaders compared to the
simple AD method. Improvements are almost negligible in the
subcritical regime. They are instead significant in both the critical
and supercritical dynamical regimes, although in the latter case there
are wide variations, with striking performance decrease for some networks. 
On average, we register
improvements of $2-5\%$. These values are in line to those that can be
measured in the corpus of small/medium networks, thus providing
additional support to the robustness and generality of our finding.
It should be stressed that the hybrid method uses a slightly larger
amount of information than the one at disposal of the individual AD
method. This might be at the root of the observed performance
increase. As a matter of fact, linear coefficients change their value
depending on the dynamical regime, so the ranking of the nodes.  On
the other hand, the improvement in effectiveness doesn't cause
drawbacks in efficiency.  Linear coefficients of the various dynamical
regimes are given. Also, the computational complexity of estimating
numerically the critical threshold $p_c$ scales linearly with system
size. {\it De facto}, the computational complexity of the overall
hybrid method is the same as the one of the individual
methods, making it applicable to very large networks.

\section{Conclusions}
The goal of this paper was to comparatively analyze the performances
of heuristic methods aimed at the identification of influential
spreaders in networks. We focused our attention on the
spreading dynamics modeled by the independent cascade model, 
and studied a total of $16$ methods for the identification
of the influential spreaders that are being used widely in influence
maximization studies.  We performed a systematic comparison
between the various methods by means of 
extensive numerical experiments on a 
large corpus of $100$ real-world networks. 
We further drew upper- and lower-bounds for
the performance values achievable in the problem 
by using respectively results from greedy optimization and random selection.
We found that the performance of many simple heuristic methods
is not far from that of the more computationally costly greedy algorithm.
In this framework, the simplest and most effective
strategy among those already on the market
that can be used to identify top spreaders in large networks
is the adaptive degree centrality. 
The method based on adaptive degree centrality displays an overall
performance score that is $96\%$ of the upper-baseline 
value in the critical
regime of spreading, if used to select a set of top spreaders
with size equal to $5\%$ of the entire network. 
Several other methods have comparable performances to adaptive degree
centrality. 
The overlap between influential spreaders selected by heuristic methods
and by the greedy algorithm is considerably lower, but this is 
not surprising given the NP-complete nature of the optimization problem.
We finally found that a potential way to get closer 
to optimality consists in  combining different centrality metrics to 
create hybrid methods.
We found that some combinations of three metrics are able to
achieve $98\%$ of the upper-baseline value in the critical regime 
of spreading.

\acknowledgements
\c{S}E and FR acknowledge support from the National Science Foundation
(CMMI-1552487). FR acknowledges support from the US Army Research
Office (W911NF-16-1-0104).

\section*{Authors contributions}
\c{S}E, CC and FR  designed the research and wrote the manuscript. 
\c{S}E  performed the research and analyzed data.

\section*{Competing interests}
The authors declare no competing interests.

\section*{Data availability}
The datasets used in this article are all publicly available 
from the cited sources.

\section*{Contact Information}
Correspondence should be addressed to F. Radicchi (\href{mailto: filiradi@indiana.edu}{filiradi@indiana.edu}).



\end{document}